\documentclass[12pt]{article}

\usepackage{graphicx}
\usepackage{amsmath}
\usepackage{amssymb}
\usepackage{amsthm}
\usepackage{color}
\usepackage{lscape}
\usepackage{soul}
\usepackage{multirow}
\usepackage{braket, enumerate}
\usepackage[font=small]{caption}

\renewcommand{\thefootnote}{\fnsymbol{footnote}}

\begin{document}

\begin{flushright}
CYCU-HEP-15-01  \\
EPHOU-15-002
\end{flushright}

\vspace{4ex}

\begin{center}

{\LARGE\bf  Gauge extension of non-Abelian discrete flavor symmetry}

\vskip 1.4cm

{\large  
Florian Beye$^{1}$\footnote{Electronic address: fbeye@eken.phys.nagoya-u.ac.jp},
Tatsuo Kobayashi$^{2}$\footnote{Electronic address: kobayashi@particle.sci.hokudai.ac.jp}
and
Shogo Kuwakino$^{3}$\footnote{Electronic address: kuwakino@cycu.edu.tw}
}
\\
\vskip 1.0cm
{\it $^1$Department of Physics, Nagoya University, Furo-cho, Chikusa-ku, Nagoya 464-8602, Japan} \\
{\it $^2$Department of Physics, Hokkaido University, Sapporo 060-0810, Japan} \\
{\it $^3$Department of Physics, Chung-Yuan Christian University, 200, Chung-Pei Rd. Chung-Li,320, Taiwan} \\

\vskip 3pt
\vskip 1.5cm

\begin{abstract}
We investigate a gauge theory realization of non-Abelian discrete flavor symmetries and apply the gauge enhancement mechanism in heterotic orbifold models to field-theoretical model building. Several phenomenologically interesting non-Abelian discrete symmetries are realized effectively from a $U(1)$ gauge theory with a permutation symmetry. We also construct a concrete model for the lepton sector based on a $U(1)^2 \rtimes S_3$ symmetry.
\end{abstract}

\end{center}

\newpage

\setcounter{footnote}{0}
\renewcommand{\thefootnote}{\arabic{footnote}}

\newpage

\section{Introduction}
The flavor structure of quarks and leptons in the standard model is mysterious. Why are there three generations? Why are their masses hierarchically different from each other?
Why do they show the specific mixing angles? It is challenging to try to solve this flavor mystery. A flavor symmetry could play an important role in particle physics models in order to understand the flavor structure of quarks and leptons. Since the Yukawa matrices of the standard model include many parameters, flavor symmetries are useful to effectively reduce the number of parameters and to obtain some predictions for experiments. In particular, non-Abelian discrete flavor symmetries can be key ingredients to make models with a suitable flavor structure. Indeed, there are many works of flavor models utilizing various non-Abelian discrete flavor symmetries (see \cite{Altarelli:2010gt,Ishimori:2010au,King:2013eh} for reviews).

It is known that some non-Abelian discrete flavor symmetries have a stringy origin. In particular, in orbifold compactification of heterotic string theory \cite{orbifold, heteroph, Katsuki:1989bf, Kobayashi:2004ud, Buchmuller, Kim, Lebedev:2006kn, Blaszczyk:2009in, Nibbelink:2013lua} (also see a review \cite{Nilles:2008gq}), non-Abelian discrete symmetries $D_4$ and $\Delta(54)$ respectively arise from one- and two-dimensional orbifolds, $S_1/Z_2$ and $T_2/Z_3$, as discussed in \cite{Kobayashi:2006wq}\footnote{Similar non-Abelian discrete symmetries including $\Delta(27)$ can appear in intersecting/magnetized D-brane models \cite{Abe:2009vi,BerasaluceGonzalez:2012vb,Hamada:2014hpa}. See also \cite{Higaki:2005ie}.}. The non-Abelian discrete symmetries originate from a geometrical property of extra-dimensional orbifolds, the permutation symmetry of orbifold fixed points, and a string selection rule between closed strings. Phenomenological applications of string derived non-Abelian discrete symmetries to flavor models are analyzed, e.g. in \cite{Ko:2007dz}. 

Furthermore, in \cite{Beye:2014nxa}, it is argued that the non-Abelian discrete symmetries $D_4$ and $\Delta(54)$ have a gauge origin within the heterotic string theory. Namely, these symmetries are respectively enhanced to continuous gauge symmetries $U(1) \rtimes Z_2$ and $U(1)^2 \rtimes S_3$ at a symmetry enhancement point in the moduli space of orbifolds. After certain scalar fields which are associated with the K\"ahler moduli fields get vacuum expectation values, the $U(1)$ symmetries break down to Abelian discrete subgroups, and there remains a $Z_4 \rtimes Z_2 \cong D_4$ or $( Z_3 \times Z_3 ) \rtimes S_3 \cong \Delta(54)$ symmetry group, respectively. This result suggests that a non-Abelian discrete symmetry can be regarded as a remnant of a continuous gauge symmetry.
Also, this result could provide us with a new insight on model building for flavor physics.

Various non-Abelian discrete symmetries other than $D_4$ and $\Delta(54)$ have been used in field-theoretical model building, e.g. $S_3$, $S_4$, $A_4$, $\Delta(3N^2)$, $\Delta(6N^2)$ (see \cite{Altarelli:2010gt,Ishimori:2010au,King:2013eh}). Thus, it is important to extend the stringy derivation of $D_4$ and $\Delta(54)$ from $U(1) \rtimes Z_2$ and $U(1)^2 \rtimes S_3$, by studying a field-theoretical derivation of other non-Abelian discrete flavor symmetries from $U(1)^m \rtimes S_n$ or $U(1)^m \rtimes Z_n$
(See also \cite{Abe:2010iv}). That is the purpose of this paper. Some of them may be reproduced from other types of string compactifications.

In this paper we consider an extension of the argument of the gauge origin in \cite{Beye:2014nxa} to field-theoretical model building. We show that phenomenologically interesting non-Abelian discrete symmetries can be embedded into $U(1)^m \rtimes S_n$ or $U(1)^m \rtimes Z_n$ continuous gauge theory. Spontaneous symmetry breaking of $U(1)^m$ to Abelian discrete symmetries leads to non-Abelian discrete flavor symmetries. In the next section we discuss a gauge theory realization of non-Abelian discrete symmetries. In section 3, we show a concrete lepton flavor model based on a $U(1)$ flavor symmetry. Section 4 is devoted to conclusions.

\section{Gauge extension of non-Abelian discrete symmetry}

In this section we investigate a field theoretical model building technique in which non-Abelian discrete symmetries have a continuous gauge symmetry origin. We start with a gauge theory with group structure of the form $U(1)^n \rtimes S_m$ or $U(1)^n \rtimes Z_m$. Then, by giving a suitable VEV to a scalar field, a non-Abelian discrete symmetry is realized effectively. 

\subsection{$S_3$ group}
We consider a $U(1) \rtimes Z_2$ model with the field contents as in Table \ref{Tab:FieldS3}. The action of the $Z_2$ symmetry on the $U(1)$ charge $q$ is given by
\begin{align} \label{S3Z2}
Z_2 : q \to -q.
\end{align}
By this we mean that the $U(1)$ gauge field $A_\mu$ transforms as $A_\mu \rightarrow -A_\mu$, and that the oppositely charged fields in this model transform into each other, e.g. $U_1 \leftrightarrow U_2$ and $M_1 \leftrightarrow M_2$. This implies that the kinetic (and gauge interaction) terms are invariant under the $Z_2$.

Now, we consider VEVs for fields $U_i$ obeying the relation
\begin{align}
\langle U_1 \rangle =  \langle U_2 \rangle.
\end{align}
This VEV relation maintains the original $Z_2$ permutation symmetry, 
\begin{align} \label{S31}
\left( 
\begin{array}{cc}
 0&1 \\
 1&0 \\
\end{array} 
\right),
\end{align}
but breaks the $U(1)$ group to a discrete $Z_3$ subgroup since the field $M_i$ has $U(1)$ charge $\pm 1/3$. The $Z_3$ charges are $1$ for the field $M_1$ and $2$ for the field $M_2$, so the $Z_3$ action is expressed by
\begin{align} \label{S32}
\left( 
\begin{array}{cc}
 \omega &0 \\
 0& \omega^{-1} \\
\end{array} 
\right),
\end{align} 
with the cubic root $\omega = e^{2\pi i/3}$. The combination of the two actions \eqref{S31} and \eqref{S32} gives rise to a non-Abelian discrete symmetry, which is nothing but $S_3 \cong Z_3 \rtimes Z_2 $. It turns out that $(M_1, M_2)$ forms a doublet of this $S_3$ group.

Next, we read off the $S_3$ representation of the other matter fields. First, the field $M$ can be regarded as the trivial singlet ${\bf 1}$ of the $S_3$ group. In the case of $(M'_1, M'_2)$, we see that these fields have trivial $Z_3$ charges. Then we can perform a change of basis as $\tilde{M}'_1 \equiv M'_1 + M'_2$ and $\tilde{M}'_2 \equiv M'_1 - M'_2$. In this basis, the $Z_2$ action is given by $\tilde{M}'_1 \to \tilde{M}'_1$ and  $\tilde{M}'_2 \to - \tilde{M}'_2$. Hence, $(M'_1, M'_2)$ forms a ${\bf 1} \oplus {\bf 1}'$ of the $S_3$ group. As a result, we can reproduce all irreducible representations of the $S_3$ group.

\begin{table}[]
\begin{center}
\begin{tabular}{|c|c|c|c|}
\hline
Field & $U(1)$ charge & $Z_3$ charge & $S_3$ rep. \\
\hline
\hline
$U_1, U_2$ & $+1, -1$ & $0,0$ & --- \\
\hline
$M_1, M_2$ & $+\frac{1}{3}, -\frac{1}{3}$ & $1, 2$ & ${\bf 2}$ \\
\hline
$M$ & $0$ & $0$ & ${\bf 1}$ \\
\hline
$M'_1, M'_2$ & $+1, -1$ & $0,0$ & ${\bf 1} \oplus {\bf 1}'$ \\
\hline
\end{tabular}
\caption[smallcaption]{Field contents of the $U(1) \rtimes Z_2$ model for the $S_3$ group. Besides the $U(1)$ charges, the charges under the unbroken discrete $Z_3$ subgroup of $U(1)$ are shown. Representations under the resulting $S_3$ group are also shown.
}
\label{Tab:FieldS3}
\end{center}
\end{table}

\subsection{$D_4$ group}
Now, we consider a $U(1) \rtimes Z_2$ model with the field contents as in Table \ref{Tab:FieldD4}. This model is based on a $U(1)$ symmetry and possesses an additional $Z_2$ symmetry which acts on the $U(1)$ charge as in the previous case \eqref{S3Z2}, so the fields transform as $U_1 \leftrightarrow U_2$ and $M_1 \leftrightarrow M_2$ etc. We consider the following VEV relation
\begin{align}
\langle U_1 \rangle = \langle U_2 \rangle.
\end{align}
This VEV relation maintains the original $Z_2$ permutation symmetry, 
\begin{align} \label{D41}
\left( 
\begin{array}{cc}
 0&1 \\
 1&0 \\
\end{array} 
\right),
\end{align}
but breaks the $U(1)$ group to its discrete $Z_4$ subgroup. The $Z_4$ charges for $M_1$ and $M_2$ are $1$ and $3$ respectively, hence the $Z_4$ action is written as
\begin{align} \label{D42}
\left( 
\begin{array}{cc}
 i &0 \\
 0& -i \\
\end{array} 
\right).
\end{align}
The combination of actions \eqref{D41} and \eqref{D42} leads to the non-Abelian discrete symmetry $D_4 \cong Z_4 \rtimes Z_2 $. It turns out that $(M_1, M_2)$ forms the doublet of the $D_4$ group.

Next, we read off the $D_4$ representation of the other matter fields. First, the field $M$ can be regarded as the trivial singlet ${\bf 1}_{++}$ of $D_4$. In the case of a set of fields $(M'_1, M'_2)$, we make redefinitions as $\tilde{M}'_1 \equiv M'_1 + M'_2$ and $ \tilde{M}'_2 \equiv M'_1 - M'_2$. In this basis, the $Z_2$ action acts as $\tilde{M}'_1 \to \tilde{M}'_1 $ and $ \tilde{M}'_2 \to - \tilde{M}'_2  $. Thus, $(M'_1, M'_2)$ forms a ${\bf 1}_{++} \oplus {\bf 1}_{--}$ of the $D_4$ group. 
For the fields $(N_1, N_2)$, both fields have $Z_4$ charge $2$. Then we can take a linear combination as $\tilde{N}_1 \equiv N_1 + N_2$ and $ \tilde{N}_2 \equiv N_1 - N_2$, and observe that the $Z_2$ action acts as $\tilde{N}_1 \to \tilde{N}_1 $ and $ \tilde{N}_2 \to - \tilde{N}_2$. 
Then $(\tilde{N}_1, \tilde{N}_2)$ forms ${\bf 1}_{+-} \oplus {\bf 1}_{-+}$ of the $D_4$ group. As a result, we can reproduce all irreducible representations of the $D_4$ group by a suitable field setup.

\begin{table}[]
\begin{center}
\begin{tabular}{|c|c|c|c|}
\hline
Field & $U(1)$ charge & $Z_4$ charge & $D_4$ rep. \\
\hline
\hline
$U_1, U_2$ & $+1,-1$ & $0,0$ & --- \\
\hline
$M_1, M_2$ & $+\frac{1}{4}, -\frac{1}{4}$ & $1,3$ &  ${\bf 2}$ \\
\hline
$M$ & $0$ & $0$ & ${\bf 1}_{++}$   \\
\hline
$M'_1, M'_2$ & $+1, -1$ & $0,0$ & ${\bf 1}_{++} \oplus {\bf 1}_{--}$  \\
\hline
$N_1, N_2$ & $+\frac{1}{2}, -\frac{1}{2}$ & $2,2$ &  ${\bf 1}_{+-} \oplus {\bf 1}_{-+}$ \\
\hline
\end{tabular}
\caption[smallcaption]{Field contents of the $U(1) \rtimes Z_2$ model for the $D_4$ group. Besides the $U(1)$ charges, the charges under the unbroken discrete $Z_4$ subgroup of $U(1)$ are shown. Representations under the resulting $D_4$ group are also shown.
}
\label{Tab:FieldD4}
\end{center}
\end{table}

\subsection{$S_4$ group} \label{S4Theory}
We consider a $U(1)^2 \rtimes S_3$ model with the field contents as in Table \ref{Tab:FieldS4}. This model has a gauge $U(1)^2$ symmetry and fields are characterized by two $U(1)$ charges $q_1$ and $q_2$. We define the two dimensional $U(1)^2$ charges $e_1, e_2$ and $e_3$ used in the table as
\begin{align}
e_1 \equiv ( \sqrt{2}, 0 ), \ e_2 \equiv ( - \frac{\sqrt{2}}{2}, \frac{\sqrt{6}}{2} ), \ e_3 \equiv ( - \frac{\sqrt{2}}{2}, - \frac{\sqrt{6}}{2} ).
\end{align}
The additional non-Abelian discrete $S_3$ symmetry is generated by a 120 degree rotation and a reflection on the two-dimensional $U(1)^2$ charge plane $(q_1, q_2)$ as 
\begin{align}
{\rm Rotation} &: 
\left( 
\begin{array}{c}
q_1 \\
q_2 \\
\end{array} 
\right)
\to
\left( 
\begin{array}{cc}
-\frac{1}{2} & \frac{\sqrt{3}}{2} \\
- \frac{\sqrt{3}}{2}& - \frac{1}{2} \\
\end{array} 
\right)
\left( 
\begin{array}{c}
q_1 \\
q_2 \\
\end{array} 
\right), \\
{\rm Reflection} &: 
\left( 
\begin{array}{c}
q_1 \\
q_2 \\
\end{array} 
\right)
\to
\left( 
\begin{array}{cc}
1 & 0 \\
0 & -1 \\
\end{array} 
\right)
\left( 
\begin{array}{c}
q_1 \\
q_2 \\
\end{array} 
\right).
\end{align}
The $S_3$ action permutes $e_1, e_2$ and $e_3$, which corresponds to a permutation of the fields as $U_1 \leftrightarrow U_2 \leftrightarrow U_3$ and $M_1 \leftrightarrow M_2 \leftrightarrow M_3$. We consider the VEV relation as
\begin{align} \label{S4VEV}
\langle U_1 \rangle = \langle U_2 \rangle = \langle U_3 \rangle.
\end{align}
This VEV relation maintains $S_3$, 
\begin{align} \label{S41}
&\left( 
\begin{array}{ccc}
 0& 1 &0 \\
 1 & 0&0 \\
 0& 0&1 \\
\end{array} 
\right), 
\left( 
\begin{array}{ccc}
 1& 0 &0 \\
 0 & 0&1 \\
 0& 1 &0 \\
\end{array} 
\right),
\end{align}
but breaks the $U(1)^2$ group down to a discrete $Z_2^2$ subgroup. The $Z_2$ charges $z_1$ and $z_2$ in Table \ref{Tab:FieldS4} are determined from the $U(1)^2$ charges as $z_1 = 2(q_1/\sqrt{2} - q_2/\sqrt{6} ) \pmod{2}$ and  $z_2 = 2(q_1/\sqrt{2} + q_2/\sqrt{6} ) \pmod{2}$. Then, the $Z_2^2$ action is given by
\begin{align} \label{S42}
\left( 
\begin{array}{ccc}
 -1& 0 &0 \\
 0 & -1&0 \\
 0& 0 &1 \\
\end{array} 
\right),
\left( 
\begin{array}{ccc}
 -1& 0 &0 \\
 0 & 1&0 \\
 0& 0 & -1 \\
\end{array} 
\right).
\end{align}
The combination of \eqref{S41} and \eqref{S42} gives rise to the non-Abelian discrete symmetry $S_4 \cong ( Z_2 \times Z_2 ) \rtimes S_3 $. It turns out that $( M_1, M_2, M_3 )$ forms the triplet ${\bf 3}$ of the $S_4$ group.

Next, we read off the $S_4$ representation of the other matter fields. First, the field $M$ can be regarded as the trivial singlet ${\bf 1}$ of $S_4$. 
In the case of the fields $( N_1, N_2, N_3 )$, we make redefinitions as $\tilde{N}_1 \equiv ( N_1 + N_2 + N_3 )/\sqrt{3}, \tilde{N}_2 \equiv ( N_1+ \omega N_2 + \omega^2 N_3 )/\sqrt{3}$ and $\tilde{N}_3 \equiv ( N_1+ \omega^2 N_2 + \omega N_3 )/\sqrt{3}$. In this basis, the three fields transform as the ${\bf 1} \oplus {\bf 2}$ of the $S_3$ group. After the VEV, these fields have the trivial $Z_2^2$ charge $(0,0)$, so they correspond to ${\bf 1} \oplus {\bf 2}$ of $S_4$. Note, that fields with opposite $U(1)^2$ charges $-e_i/2$ have, after $U(1)^2$ breaking, the same $Z_2^2$ charges as the fields $M_i$. Hence, such fields also lead to the ${\bf 3}$ of $S_4$. As a result, we can realize the ${\bf 1}, {\bf 1} \oplus {\bf 2}, {\bf 3}$ representations of the $S_4$ group in this setup.

\begin{table}[]
\begin{center}
\begin{tabular}{|c|c|c|c|}
\hline
Field & $U(1)^2$ charge & $Z_2^2$ charge & $S_4$ rep. \\
\hline
\hline
$U_1, U_2, U_3$ & $- e_1, - e_2, - e_3 $ & $(0,0), (0,0), (0,0) $ & --- \\
\hline
$M_1, M_2, M_3$ & $\frac{e_1}{2}, \frac{e_2}{2}, \frac{e_3}{2} $ & $(1, 1), (1, 0), (0,  1)$ & ${\bf 3}$ \\
\hline
$M$ & $0$ & $(0, 0)$ & ${\bf 1}$ \\
\hline
$N_1, N_2, N_3$ & $e_1, e_2, e_3$ & $(0, 0), (0,0), (0,0)$ & ${\bf 1} \oplus {\bf 2} $ \\
\hline
\end{tabular}
\caption[smallcaption]{
Field contents of the $U(1)^2 \rtimes S_3$ model for the $S_4$ group. Besides the $U(1)^2$ charges, the charges under the unbroken discrete $Z_2^2$ subgroup of $U(1)^2$ are shown. Representations under the resulting $S_4$ group are also shown.
}
\label{Tab:FieldS4}
\end{center}
\end{table}

We have introduced the specific combination of $U(1)^2$ charges, $e_1$, $e_2$, and $e_3$ which can be interpreted as weights of the fundamental $SU(3)$ triplet (or anti-triplet) representation. Then, the action of the $S_3$ group on the $e_i$ corresponds to the action of the Weyl group of $SU(3)$ on the triplet weights. Thus, one might wonder about a $SU(3)$ origin of this setup. In fact, $U(1)^2 \rtimes S_3$ is a subgroup of $SU(3)$ where $U(1)^2$ furnishes maximal torus and $S_3$ is a lift of the Weyl group into $SU(3)$. Also, note that the representation matrices \eqref{S41} of $S_3$ do not actually belong to $SU(3)$, so they give rise to genuine $U(1)^2 \rtimes S_3$ representations. The fundamental triplet and anti-triplet of $SU(3)$ also give rise to $U(1)^2 \rtimes S_3$ representations which we did not cover here (in these cases, the representation matrices are given by those in \eqref{S41} amended by a minus sign). For a short remark on these kinds of representations please refer to the conclusion section.

Furthermore, in a stringy realization of $\Delta(54)$, the $SU(3)$ gauge symmetry appears in toroidal compactification at a symmetry enhanced point. Then, by a $Z_3$ orbifolding the charged root vectors are projected out \cite{Beye:2014nxa}, leaving a symmetry group $U(1)^2 \rtimes S_3$. 

To realize $\Delta(54)$, $A_4$ and $\Delta(27)$ in the next subsections, we also use the vectors $e_1$, $e_2$ and $e_3$, as well as the Weyl reflections and the Coxeter elements.

\subsection{$\Delta(54)$ group} \label{Delta54}
We consider a $U(1)^2 \rtimes S_3$ model for the $\Delta(54)$ group, with field contents given as in Table \ref{Tab:FieldDelta54}. The difference from the previous subsection is that the matter fields now have relative $U(1)$ charges of $1/3$ when compared to the fields $U_i$. Then, by the VEV relation \eqref{S4VEV} for the field $U_i$, the $S_3$ symmetry remains but $U(1)^2$ is broken down to its Abelian subgroup $Z_3^2$. The two $Z_3$ charges $z_1, z_2$ in Table \ref{Tab:FieldDelta54} are determined as $z_1 = 3(q_1/\sqrt{2} - q_2/\sqrt{6} ) \pmod{3}$ and  $z_2 = 3(q_1/\sqrt{2} + q_2/\sqrt{6} ) \pmod{3}$, and the $Z_3^2$ action is described by
\begin{align} \label{Delta541}
\left( 
\begin{array}{ccc}
 \omega & 0 &0 \\
 0 & \omega^{-1} &0 \\
 0& 0 &1 \\
\end{array} 
\right),
\left( 
\begin{array}{ccc}
 \omega & 0 &0 \\
 0 & 1 &0 \\
 0& 0 & \omega^{-1}  \\
\end{array} 
\right).
\end{align}
The actions \eqref{S41} and \eqref{Delta541} together generate the non-Abelian discrete symmetry $\Delta(54) \cong ( Z_3 \times Z_3 ) \rtimes S_3 $. It turns out that $(M_1, M_2, M_3)$ forms the triplet ${\bf 3}_{1(1)}$ of the $\Delta(54)$ group.

Next, we read off the representation of the other matter fields under the $\Delta(54)$ group. First, the fields $(M'_1, M'_2, M'_3)$, which have opposite $U(1)^2$ charges and $Z_3^2$ charges when compared to the $M_i$ field, lead to the ${\bf 3}_{1(2)}$ of $\Delta(54)$. The field $M$ can be regarded as the trivial singlet ${\bf 1}_+$ of $\Delta(54)$. In the case of the fields $(N_1, N_2, N_3)$, we use the linear combinations $\tilde{N}_1 \equiv ( N_1 + N_2 + N_3 )/\sqrt{3}, \tilde{N}_2 \equiv ( N_1+ \omega N_2 + \omega^2 N_3 )/\sqrt{3}$ and $ \tilde{N}_3 \equiv ( N_1+ \omega^2 N_2 + \omega N_3 )/\sqrt{3}$. In this basis, one sees that they transform as a ${\bf 1} \oplus {\bf 2}$ of the $S_3$ group. After the VEV, these fields have trivial $Z_3^2$ charges, so they correspond to ${\bf 1}_+ \oplus {\bf 2}_1$ of the $\Delta(54)$ group. Note, that instead of the $M_i$ which have $U(1)^2$ charges $e_i/3$, we can also introduce fields with charges $-2e_i/3$. Since the $Z_3^2$ charges of such fields are identical to the $M_i$, they also lead to the ${\bf 3}_{1(1)}$ representation. As the result, we can realize ${\bf 1}_+, {\bf 1}_+ \oplus {\bf 2}_1, {\bf 3}_{1(1)}, {\bf 3}_{1(2)}$ representations of the $\Delta(54)$ group in our setup.

\begin{table}[]
\begin{center}
\begin{tabular}{|c|c|c|c|}
\hline
Field & $U(1)^2$ charge & $Z_3^2$ charge & $\Delta(54)$ rep. \\
\hline
\hline
$U_1, U_2, U_3$ & $- e_1, - e_2, - e_3 $ & $(0,0), (0,0), (0,0) $ & --- \\
\hline
$M_1, M_2, M_3$ & $\frac{e_1}{3}, \frac{e_2}{3}, \frac{e_3}{3}$ & $(1, 1), (2, 0), (0,  2 )$ & ${\bf 3}_{1(1)}$ \\
\hline
$M'_1, M'_2, M'_3$ & $ - \frac{e_1}{3}, - \frac{e_2}{3}, - \frac{e_3}{3}$ & $(2, 2), (1, 0), (0, 1)$ & ${\bf 3}_{1(2)} $ \\
\hline
$M$ & $0$ & $(0, 0)$ & ${\bf 1}_+$ \\
\hline
$N_1, N_2, N_3$ & $e_1, e_2, e_3$ & $(0, 0), (0,0), (0,0)$ & ${\bf 1}_+ \oplus {\bf 2}_1 $ \\
\hline
\end{tabular}
\caption[smallcaption]{
Field contents of the $U(1)^2 \rtimes S_3$ model for the $\Delta(54)$ group. Besides the $U(1)^2$ charges, the charges under the unbroken discrete $Z_3^2$ subgroup of $U(1)^2$ are shown. Representations under the resulting $\Delta(54)$ group are also shown.
}
\label{Tab:FieldDelta54}
\end{center}
\end{table}

\subsection{$A_4$ group}
We consider a $U(1)^2 \rtimes Z_3$ model with the field contents as in Table \ref{Tab:FieldA42}. There, we add fields $A_i$ to the field contents of the model for the $S_4$ group (Table \ref{Tab:FieldS4}). We define the two-dimensional $U(1)^2$ charges as
\begin{align}
w_1 &\equiv ( \frac{\sqrt{2}}{2}, \frac{\sqrt{6}}{6} ), \ 
w_2 \equiv ( - \frac{\sqrt{2}}{2}, \frac{\sqrt{6}}{6} ), \ 
w_3 \equiv ( 0, - \frac{\sqrt{6}}{3} ).
\end{align}
The introduction of $A_i$ fields breaks the original $S_3$ symmetry to a $Z_3$ symmetry (under reflections, their $U(1)^2$ charges are not mapped onto each other). Then, this model has a $U(1)^2 \rtimes Z_3$ structure, the $Z_3$ symmetry acting as $U_1 \rightarrow U_2 \rightarrow U_3 \rightarrow U_1$ and $M_1 \rightarrow M_2 \rightarrow M_3 \rightarrow M_1$, etc. We consider a VEV relation as
\begin{align} \label{VEV}
\langle U_1 \rangle =  \langle U_2 \rangle =  \langle U_3 \rangle.
\end{align}
This VEV relation maintains $Z_3$, 
\begin{align} \label{A41}
&\left( 
\begin{array}{ccc}
 0& 0 &1 \\
 1 & 0&0 \\
 0& 1&0 \\
\end{array} 
\right),
\end{align}
but breaks $U(1)^2$ to its Abelian subgroup $Z_2^2$. The two $Z_2$ charges $z_1, z_2$ in Table \ref{Tab:FieldA42} are determined by $z_1 = 2(q_1/\sqrt{2} - q_2/\sqrt{6} ) \pmod{2}$ and  $z_2 = 2(q_1/\sqrt{2} + q_2/\sqrt{6} ) \pmod{2}$,  and the $Z_2^2$ action is given by
\begin{align} \label{A42}
\left( 
\begin{array}{ccc}
 -1& 0 &0 \\
 0 & -1&0 \\
 0& 0 &1 \\
\end{array} 
\right),
\left( 
\begin{array}{ccc}
 -1& 0 &0 \\
 0 & 1&0 \\
 0& 0 & -1 \\
\end{array} 
\right).
\end{align}
By combining \eqref{A41} and \eqref{A42}, this leads to non-Abelian discrete symmetry $A_4 \cong ( Z_2 \times Z_2 ) \rtimes Z_3 $. It turns out that $(M_1, M_2, M_3)$ forms the triplet ${\bf 3}$ of the $A_4$ group.

Next, we read off the $A_4$ representation of the other fields. First the field $M$ can be regarded as the trivial singlet ${\bf 1}$ of $A_4$. The fields $(A_1, A_2, A_3)$ have a similar structure to the fields $M_i$, and they also lead to a ${\bf 3}$ of $A_4$.
In the case of the fields $(N_1, N_2, N_3)$, we use the linear combinations $\tilde{N}_1 \equiv ( N_1 + N_2 + N_3 )/\sqrt{3}, \tilde{N}_2 \equiv ( N_1+ \omega N_2 + \omega^2 N_3 )/\sqrt{3}$ and $ \tilde{N}_3 \equiv ( N_1+ \omega^2 N_2 + \omega N_3 )/\sqrt{3}$. In this basis, the three fields transform as ${\bf 1} \oplus {\bf 1}' \oplus {\bf 1}''$ of $Z_3$. After the VEV, these fields have trivial $Z_2^2$ charges, so they correspond to ${\bf 1} \oplus {\bf 1}' \oplus {\bf 1}''$ of the $A_4$ group. Note that other fields $(M'_1, M'_2, M'_3)$ with $U(1)^2$ charges $(2n + 1) e_i/2$, where $n$ is an integer, also lead to ${\bf 3}$ representation since they have same $Z_2^2$ charges as $M_i$. As a result, we can realize ${\bf 1}, {\bf 1} \oplus {\bf 1}' \oplus {\bf 1}'', {\bf 3}$ representations of $A_4$ in this setup. 

\begin{table}[]
\begin{center}
\begin{tabular}{|c|c|c|c|}
\hline
Field & $U(1)^2$ charge & $Z_2^2$ charge & $A_4$ rep. \\
\hline
\hline
$U_1, U_2, U_3$ & $- e_1, - e_2, - e_3 $ & $(0,0), (0,0), (0,0)$ & --- \\
\hline
$M_1, M_2, M_3$ & $\frac{e_1}{2}, \frac{e_2}{2}, \frac{e_3}{2}$ & $(1, 1), (1, 0), (0,  1)$ & ${\bf 3}$ \\
\hline
$M$ & $0$ & $(0, 0)$ & ${\bf 1}$ \\
\hline
$N_1, N_2, N_3$ & $e_1, e_2, e_3$ & $(0, 0), (0, 0), (0, 0)$ & ${\bf 1} \oplus {\bf 1}' \oplus {\bf 1}''$ \\
\hline
$A_1, A_2, A_3$ & $ \frac{3 w_1 }{2}, \frac{3 w_2 }{2}, \frac{3 w_3 }{2}  $ & $(1, 0), (0, 1), (1, 1) $ & ${\bf 3}$ \\
\hline
\end{tabular}
\caption[smallcaption]{
Field contents of the $U(1)^2 \rtimes Z_3$ model for the $A_4$ group. Besides the $U(1)^2$ charges, the charges under the unbroken discrete $Z_2^2$ subgroup of $U(1)^2$ are shown. Representations under the resulting $A_4$ group are also shown.
}
\label{Tab:FieldA42}
\end{center}
\end{table}

\subsection{$\Delta(27)$ group}
We consider a $U(1)^2 \rtimes Z_3$ model with the field contents as in Table \ref{Tab:FieldDelta27}. There, we have added fields $A_i$ and $B_i$ to the field content of the $\Delta(54)$ model (Table \ref{Tab:FieldDelta54}). These fields break the $S_3$ symmetry to a $Z_3$ symmetry. We now consider the VEV relation \eqref{VEV}, which maintains $Z_3$ \eqref{A41} but breaks $U(1)^2$ to its Abelian subgroup $Z_3^2$. The two $Z_3$ charges $z_1, z_2$ in Table \ref{Tab:FieldDelta27} are determined as $z_1 = 3(q_1/\sqrt{2} - q_2/\sqrt{6} ) \pmod{3}$ and  $z_2 = 3(q_1/\sqrt{2} + q_2/\sqrt{6} ) \pmod{3}$. Also, the $Z_3^2$ action is given by
\begin{align} \label{Delta272}
\left( 
\begin{array}{ccc}
 \omega & 0 &0 \\
 0 & \omega^{-1} &0 \\
 0& 0 &1 \\
\end{array} 
\right),
\left( 
\begin{array}{ccc}
 \omega & 0 &0 \\
 0 & 1 &0 \\
 0& 0 & \omega^{-1}  \\
\end{array} 
\right).
\end{align}
The generators \eqref{A41} and \eqref{Delta272} generate a non-Abelian discrete symmetry $\Delta(27) \cong ( Z_3 \times Z_3 ) \rtimes Z_3 $. It turns out that $(M_1, M_2, M_3)$ forms the triplet ${\bf 3}_{[0][1]}$ of the $\Delta(27)$ group. 

Next we read off the representation of the other matter fields under the $\Delta(27)$ group. First, the fields $(M'_1, M'_2, M'_3)$ which have opposite $U(1)^2$ charges and $Z_3^2$ charges when compared to the fields $M_i$ lead to a ${\bf 3}_{[0][2]}$ of the $\Delta(27)$ group. The field $M$ can be regarded as the trivial singlet ${\bf 1}_{0,0}$ of $\Delta(27)$. In the case of the fields $(N_1, N_2, N_3)$, we use the linear combinations $\tilde{N}_1 \equiv ( N_1 + N_2 + N_3 )/\sqrt{3}, \tilde{N}_2 \equiv ( N_1+ \omega N_2 + \omega^2 N_3 )/\sqrt{3}$ and $ \tilde{N}_3 \equiv ( N_1+ \omega^2 N_2 + \omega N_3 )/\sqrt{3}$. In this basis, the three fields transform as ${\bf 1} \oplus {\bf 1}' \oplus {\bf 1}''$ of $Z_3$. After the VEV, these fields have the trivial $Z_3^2$ charges, so they correspond to ${\bf 1}_{0,0} \oplus {\bf 1}_{1,0} \oplus {\bf 1}_{2,0} $ of the $\Delta(27)$ group. Next we consider the fields $(A_1, A_2, A_3)$. They have degenerate $Z_3^2$ charges, so by diagonalization we observe that they transform as ${\bf 1} \oplus {\bf 1}' \oplus {\bf 1}''$ under $Z_3$. Then, these fields lead to a ${\bf 1}_{0,2} \oplus {\bf 1}_{1,2} \oplus {\bf 1}_{2,2} $ of the $\Delta(27)$ group. Similarly $(B_1, B_2, B_3)$ lead to ${\bf 1}_{0,1} \oplus {\bf 1}_{1,1} \oplus {\bf 1}_{2,1}$ of the $\Delta(27)$ group. As a result, we can realize the ${\bf 1}_{0,0}, {\bf 1}_{0,0} \oplus {\bf 1}_{1,0} \oplus {\bf 1}_{2,0}, {\bf 1}_{0,1} \oplus {\bf 1}_{1,1} \oplus {\bf 1}_{2,1}, {\bf 1}_{0,2} \oplus {\bf 1}_{1,2} \oplus {\bf 1}_{2,2},  {\bf 3}_{[0][1]}, {\bf 3}_{[0][2]}$  representations of the $\Delta(27)$ group in this setup.

\begin{table}[]
\begin{center}
\begin{tabular}{|c|c|c|c|}
\hline
Field & $U(1)^2$ charge & $Z_3^2$ charge & $\Delta(27)$ rep. \\
\hline
\hline
$U_1, U_2, U_3$ & $- e_1, - e_2, - e_3 $ & $(0,0), (0,0), (0,0)$ & --- \\
\hline
$M_1, M_2, M_3$ & $\frac{e_1}{3}, \frac{e_2}{3}, \frac{e_3}{3}$ & $(1, 1), ( 2, 0), (0, 2)$ & ${\bf 3}_{[0][1]}$ \\
\hline
$M'_1, M'_2, M'_3 $ & $ - \frac{e_1}{3}, - \frac{e_2}{3}, - \frac{e_3}{3}$ & $(2, 2), (1, 0), (0, 1)$ & ${\bf 3}_{[0][2]} $ \\
\hline
$M$ & $0$ & $(0, 0)$ & ${\bf 1}_{0,0}$ \\
\hline
$N_1, N_2, N_3$ & $e_1, e_2, e_3$ & $(0, 0), (0,0), (0,0)$ & ${\bf 1}_{0,0} \oplus {\bf 1}_{1,0} \oplus {\bf 1}_{2,0} $ \\
\hline
$A_1, A_2, A_3$ & $ w_1, w_2, w_3  $ & $(1, 2), (1, 2), (1, 2) $ & ${\bf 1}_{0,2} \oplus {\bf 1}_{1,2} \oplus {\bf 1}_{2,2} $ \\
\hline
$B_1, B_2, B_3$ & $ 2 w_1, 2 w_2, 2 w_3  $ & $ (2, 1), (2, 1), (2, 1) $ & ${\bf 1}_{0,1} \oplus {\bf 1}_{1,1} \oplus {\bf 1}_{2,1} $ \\
\hline
\end{tabular}
\caption[smallcaption]{
Field contents of the $U(1)^2 \rtimes Z_3$ model for the $\Delta(27)$ group. Besides the $U(1)^2$ charges, the charges under the unbroken discrete $Z_3^2$ subgroup of $U(1)^2$ are shown. Representations under the resulting $\Delta(27)$ group are also shown.
}
\label{Tab:FieldDelta27}
\end{center}
\end{table}

\section{$U(1)^2 \rtimes S_3$ lepton flavor model}
In this section we present a concrete model for the lepton sector based on the $U(1)^2 \rtimes S_3$ symmetry, which is related to the $\Delta(54)$ discrete symmetry discussed in Section \ref{Delta54}. Several interesting flavor models based on the $\Delta(54)$ symmetry have been investigated in \cite{Lam:2008sh, Escobar:2008vc, Ishimori:2008uc, King:2009ap, Escobar:2011mq}. 

Here we consider a supersymmetric model with $U(1)^2 \rtimes S_3 \times Z_2$ symmetry, and with the field content as in Table \ref{Tab:PU1Model}. There, in addition to the MSSM fields (the lepton doublets $(L_e, L_{\mu}, L_{\tau})$, the right-handed lepton fields $(e^c, \mu^c, \tau^c)$ and Higgs doublet pairs $(H_u, H_d)$) we introduce flavon fields $A_i, B_i, C_i$ and $D_i$. The VEV of the flavon fields breaks the $U(1)^2 \rtimes S_3$ symmetry completely. Corresponding representations under $\Delta(54)$ are also shown in Table \ref{Tab:PU1Model}. It is also possible to add other flavon fields, e.g. fields $U_i$ in Table \ref{Tab:FieldDelta54}, and consider the situation where the VEV of the fields, $\langle U_1 \rangle = \langle U_2 \rangle = \langle U_3 \rangle$, breaks the symmetry as $U(1)^2 \rtimes S_3 \to \Delta(54)$ at an intermediate scale. In this paper we do not consider this possibility.

\begin{table}[]
\begin{center}
\begin{tabular}{|c|c|c|c|}
\hline
Field &  $U(1)^2$ charge & $ Z_2 $ charge & $\Delta(54)$ rep. \\
\hline
\hline
$( L_e, L_\mu, L_\tau )$ &  $ ( \frac{2 e_1}{3}, \frac{2 e_2}{3}, \frac{2 e_3}{3} ) $ & $ 0 $ & ${\bf 3}_{1(2)} $ \\
\hline
$( e^c, \mu^c, \tau^c )$ &  $ ( -3e_1, -3e_2, -3e_3 ) $ & $ 1 $ & ${\bf 1}_+ \oplus {\bf 2}_1 $ \\
\hline
$H_{u}$ &  $0$ & $ 0 $ & $ {\bf 1}_+ $ \\
\hline
$H_{d}$ &  $0$ & $ 0 $ & $ {\bf 1}_+ $ \\
\hline
$( A_1, A_2, A_3 )$ &  $( \frac{2 e_1}{3}, \frac{2 e_2}{3}, \frac{2 e_3}{3} )$ & $ 0 $ & $ {\bf 3}_{1(2)} $ \\
\hline
$( B_1, B_2, B_3 )$ &  $( - \frac{4 e_1}{3}, - \frac{4 e_2}{3}, - \frac{4 e_3}{3} )$  & $ 0 $ & $ {\bf 3}_{1(2)} $ \\
\hline
$( C_1, C_2, C_3 )$ &  $( \frac{e_1}{3}, \frac{e_2}{3}, \frac{e_3}{3} )$ & $ 0 $ & $ {\bf 3}_{1(1)} $ \\
\hline
$( D_1, D_2, D_3 )$ &  $(  \frac{7 e_1}{3},  \frac{7 e_2}{3},  \frac{7 e_3}{3} )$ & $ 1 $ & $ {\bf 3}_{1(1)} $ \\
\hline
\end{tabular}
\caption[smallcaption]{
Field contents of the $U(1)^2 \rtimes S_3 \times Z_2$ lepton flavor model. $U(1)^2$ charges and $Z_2$ charges are shown. Representations under the $\Delta(54)$ group are also shown.
}
\label{Tab:PU1Model}
\end{center}
\end{table}

\subsection{Yukawa mass matrices}
First, we consider the Yukawa sector of the model. By invariance under $U(1)^2 \rtimes S_3 \times Z_2$, the superpotentials of the neutrino sector and the charged lepton sector are given by
\begin{align}
W_\nu &= 
y^\nu_1 \left( B_1 L_e  L_e +  B_2 L_\mu  L_\mu +  B_3 L_\tau  L_\tau \right)  H_u H_u/\Lambda^2 \nonumber \\
&+ y^\nu_2 \left( A_1 ( L_\mu  L_\tau + L_\tau  L_\mu ) +  A_2 ( L_e  L_\tau + L_\tau  L_e ) +
 A_3 ( L_e  L_\mu + L_\mu  L_e )  \right)  H_u H_u/\Lambda^2 \nonumber \\
&+ y^\nu_3 \left( C_1^2 ( L_\mu  L_\tau + L_\tau  L_\mu ) +  C_2^2 ( L_e  L_\tau + L_\tau  L_e ) +
 C_3^2 ( L_e  L_\mu + L_\mu  L_e )  \right)  H_u H_u/\Lambda^3,
\end{align}
and
\begin{align}
W_e &= 
y^e_1 \left( D_1 L_e e^c +  D_2 L_\mu \mu^c +  D_3 L_\tau \tau^c \right)  H_d/\Lambda,
\end{align}
respectively. Here, we assume a UV cutoff scale $\Lambda$. Then the mass matrices are given by
\begin{align}  \label{MMForPU1}
M_\nu &= \frac{v_u^2}{\Lambda^2} 
\left(
\begin{array}{ccc}
y^\nu_1 b_1 & y^\nu_2 a_3  & y^\nu_2 a_2 \\
 y^\nu_2 a_3 & y^\nu_1 b_2  & y^\nu_2 a_1 \\
 y^\nu_2 a_2 & y^\nu_2 a_1 & y^\nu_1 b_3  \\
\end{array} 
\right)
+
\frac{y^\nu_3 v_u^2}{\Lambda^3} 
\left(
\begin{array}{ccc}
0 &  c_3^2  &  c_2^2 \\
  c_3^2 & 0  &  c_1^2 \\
  c_2^2 &  c_1^2 & 0  \\
\end{array} 
\right), \\
M_e &= \frac{y^e_1 v_d}{\Lambda} 
\left(
\begin{array}{ccc}
d_1 & 0  & 0 \\
 0 & d_2  & 0 \\
 0 & 0 & d_3  \\
\end{array} 
\right),
\label{CLForU1E}
\end{align}
where we used the following definition for the VEVs of the flavon fields: 
\begin{align}
\langle ( A_1, A_2, A_3 ) \rangle &= ( a_1, a_2, a_3 ), \\
\langle ( B_1, B_2, B_3 ) \rangle &= ( b_1, b_2, b_3 ), \\
\langle ( C_1, C_2, C_3 ) \rangle &= ( c_1, c_2, c_3 ). \\
\langle ( D_1, D_2, D_3 ) \rangle &= ( d_1, d_2, d_3 ).
\end{align}
Note that the charged lepton mass matrix is diagonal.
Thus, the mixing angles are determined only by the neutrino mass matrix.

\subsection{Flavon potential and vacuum alignment}
Next we consider the flavon sector. The superpotential up to three-point level including only flavon fields is given by
\begin{align} \label{FSP}
	W_f & = \lambda_1 A_1 A_2 A_3 + \lambda_2 B_1 B_2 B_3 + \lambda_3 C_1 C_2 C_3 
		 + \lambda_4 \left(A_1^2 B_1 + A_2^2 B_2 + A_3^2 B_3\right)  .
\end{align}
The F-flatness condition for the flavon superpotential leads to (for $i \neq j \neq k \neq i$) 
\begin{align}
\label{e:fA} 0 &= \frac{\partial W_f}{\partial{A_k}} = \lambda_1 A_i A_j + 2 \lambda_4 A_k B_k,\\
\label{e:fB} 0 &= \frac{\partial W_f}{\partial{B_k}} = \lambda_2 B_i B_j + \lambda_4 A_k^2,\\
\label{e:fC} 0 &= \frac{\partial W_f}{\partial{C_k}} = \lambda_3 C_i C_j,\\
\label{e:fD} 0 &= \frac{\partial W_f}{\partial{D_k}} = 0.
\end{align}
There are two branches of solutions:
\begin{enumerate}[(a)]
\item Let us first assume $A_i \neq 0$ and $B_i \neq 0$. Then we can solve \eqref{e:fA} for $B_k$ and insert the solution in to \eqref{e:fB}. Then, we obtain the condition $4\lambda_4^3 = -\lambda_2\lambda_1^2$, so we can choose the VEVs as:
\begin{align}\label{e:VEV1}
	\left\langle A_i \right\rangle = \begin{pmatrix}
		a_1 \\ a_2 \\ a_3
	\end{pmatrix}, \;\;\; \left\langle B_i \right\rangle = -\frac{\lambda_1}{2\lambda_4}\begin{pmatrix} \frac{a_2 a_3}{a_1} \\ \frac{a_3 a_1}{a_2}  \\ \frac{a_1 a_2}{a_3} \end{pmatrix}.
\end{align}
\item If not all $A_i \neq 0$ or $B_i \neq 0$ then there exist solutions, and they can be brought into the following form by an $S_3$ transformation:
\begin{align}\label{e:VEV2}
	\left\langle A_i \right\rangle = \begin{pmatrix}
		0 \\ 0 \\ a_3
	\end{pmatrix}, \;\;\; \left\langle B_i \right\rangle = \begin{pmatrix} b_1 \\ b_2  \\ 0 \end{pmatrix},
\end{align}
with the condition $\lambda_2 b_1 b_2 + \lambda_4 a_3^2 = 0$.
\end{enumerate}
Furthermore, the VEVs of any two components $C_i$ must be zero. In the following we assume 
\begin{align}
	\left\langle C_i \right\rangle = \begin{pmatrix}
		c_1 \\ 0 \\ 0
	\end{pmatrix}.
\end{align}
The $D_i$ are not constrained from F-flatness.

\subsection{Neutrino mass/mixing properties}
In the following we consider only the case (a). By inserting the VEVs the mass matrix becomes 
\begin{align} \label{MMEPU1}
M_\nu = \frac{v_u^2}{\Lambda^2} 
\left(
\begin{array}{ccc}
- y^\nu_1 \frac{ \lambda_1 }{ 2 \lambda_4 } \frac{a_2 a_3}{a_1} & y^\nu_2 a_3  & y^\nu_2 a_2 \\
 y^\nu_2 a_3 & - y^\nu_1 \frac{ \lambda_1 }{ 2 \lambda_4 } \frac{a_1 a_3}{a_2}  & y^\nu_2 a_1 \\
 y^\nu_2 a_2 & y^\nu_2 a_1 & - y^\nu_1 \frac{ \lambda_1 }{ 2 \lambda_4 } \frac{a_1 a_2}{a_3}  \\
\end{array} 
\right)
+
\frac{y^\nu_3 v_u^2}{\Lambda^3} 
\left(
\begin{array}{ccc}
0 &  0  & 0 \\
  0 & 0  & c_1^2 \\
 0 & c_1^2 & 0  \\
\end{array} 
\right).
\end{align}
For the later convenience we define the following parameters 
\begin{align}
a'_2 \equiv \frac{a_2}{a_1}, \ 
a'_3 \equiv \frac{a_3}{a_1}, \ 
A \equiv \frac{v_u^2 y^\nu_2 a_1 }{\Lambda^2}, \
B \equiv - \frac{y_1^\nu}{y_2^\nu} \frac{ \lambda_1 }{ 2\lambda_4 }, \
C \equiv \frac{y_3^\nu}{y_2^\nu} \frac{ c_1^2}{a_1 \Lambda},
\end{align}
($A$, $B$ and $C$ not to be confused with the flavon fields $A_i$, $B_i$ and $C_i$) and rewrite the mass matrix \eqref{MMEPU1} as
\begin{align} 
\label{MMN2}
M_\nu = 
A
\left(
\begin{array}{ccc}
B a'_2 a'_3  &  a'_3   & a'_2 \\
 a'_3   & B \frac{a'_3}{a'_2}  & 1 + C \\
 a'_2 & 1 + C & B \frac{a'_2}{a'_3}  \\
\end{array} 
\right).
\end{align}
It turns out that this mass matrix has the following relations,
\begin{align} 
\label{ratioE1} \frac{M_{22}}{M_{11}} &= \left( \frac{M_{23}  -AC }{M_{13}} \right)^2,  \\
\label{ratioE2} \frac{M_{33}}{M_{22}} &= \left( \frac{M_{13}}{M_{12} } \right)^2,  \\
\frac{M_{11}}{M_{33}} &= \left( \frac{M_{12} }{M_{23} - AC } \right)^2.
\end{align}
Note that the three equations are dependent. Actually, the third equation is a consequence of the first and the second equations. The first equation \eqref{ratioE1} can be solved by $AC$ as
\begin{align}
AC = M_{23} \pm M_{13} \sqrt{ \frac{ M_{22} }{ M_{11} }  },
\end{align}
thus if the mass matrix $M_{\nu}$ is fixed, the parameter $AC$ can be derived. Hence, \eqref{ratioE2} is a prediction for ratios of elements of the neutrino mass matrix $M_{\nu}$. 

\begin{figure}[]
  \begin{center}
    \includegraphics[clip,width=10.0cm]{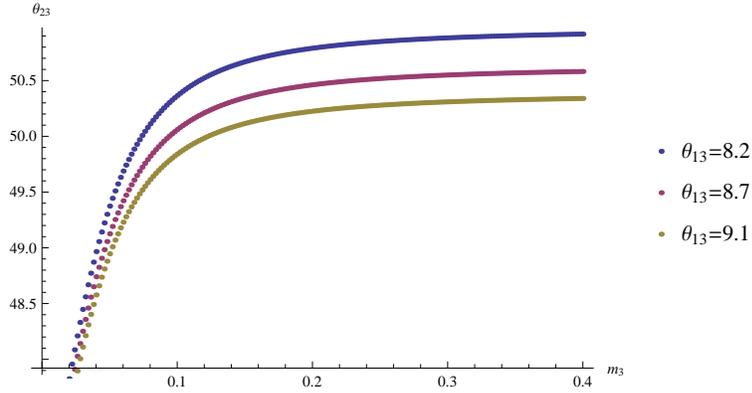}
    \caption{The mixing angle $\theta_{23}$ (in degrees) against the third generation neutrino mass $m_3$ (in eV) for various values of the mixing angle $\theta_{13}$ (in degrees).}
    \label{fig:tablem3t23}
  \end{center}
\end{figure}

Now, we investigate whether this model can explain the experimental values of mass hierarchies and mixings. In our model, the charged lepton mass matrix \eqref{CLForU1E} already takes a diagonal form, so the PMNS mixing matrix $U_{\rm PMNS}$ is given by a unitary matrix $U_{\nu}$ which diagonalizes the neutrino mass matrix \eqref{MMN2} as
\begin{align}
\label{PMNS} U_{{\rm PMNS}} &= U_{\nu} = R_{23} U_{13} R_{12} P_{12}.
\end{align}
Here, the rotation matrices are defined by three mixing angles $(\theta_{12}, \theta_{23}, \theta_{13})$ and three CP phases $(\delta, \beta_1, \beta_2)$ as
\begin{align} \label{MM}
R_{23} &=  
\left(
\begin{array}{ccc}
1 & 0  & 0 \\
 0 & \cos{\theta_{23}}  & \sin{\theta_{23}} \\
 0 & - \sin{\theta_{23}} & \cos{\theta_{23}}  \\
\end{array} 
\right), 
\ 
U_{13} =  
\left(
\begin{array}{ccc}
\cos{\theta_{13}} & 0  & \sin{\theta_{13}} e^{-i\delta} \\
 0 & 1  & 0 \\
 - \sin{\theta_{13}} e^{i \delta } & 0 & \cos{\theta_{13}}  \\
\end{array} 
\right), \\
R_{12} &=  
\left(
\begin{array}{ccc}
\cos{\theta_{12}} & \sin{\theta_{12}}  & 0 \\
 - \sin{\theta_{12}} & \cos{\theta_{12}}  & 0 \\
 0 & 0 & 1  \\
\end{array} 
\right),
\ 
P_{12} =  
\left(
\begin{array}{ccc}
e^{i\beta_1} & 0  & 0 \\
 0 & e^{i\beta_2}  & 0 \\
 0 & 0 & 1  \\
\end{array} 
\right).
\end{align}
For simplicity, here we consider only the case where 
\begin{align}
\delta &= \beta_1 = \beta_2 = 0.
\end{align}
We also set the mixing angle $\theta_{12}$ to fit the experimental value as
\begin{align}
\theta_{12} &= 35.3^\circ. 
\end{align}
Then the mixing matrix \eqref{PMNS} is a real matrix. As for the neutrino mass differences, we wish to reproduce the case of the inverted hierarchy:
\begin{align}
\Delta m_{21}^2 &= m_2^2 - m_1^2 = 7.60 \times 10^{-5} \ {\rm eV}^2, \\
\Delta m_{31}^2 &= m_3^2 - m_1^2 = - 2.38 \times 10^{-3} \ {\rm eV}^2,
\end{align}
and regard the third family neutrino mass $m_3$ as a parameter. These values are consistent with the global analysis in \cite{Forero:2014bxa} within $2\sigma$ range. The neutrino mass matrix is then obtained as
\begin{align}
M_\nu &= U_{{\rm PMNS}}\ M U_{{\rm PMNS}}^{T},
\end{align}
where $M = {\rm diag} ( m_1, m_2, m_3 )$. In Figure \ref{fig:tablem3t23}, we show a prediction for various values of $(m_3, \theta_{13}, \theta_{23})$ from the ratio condition of this mass matrix \eqref{ratioE2}. In the figure we show solutions of the mixing angle $\theta_{23}$ against the third generation neutrino mass $m_3$ for \eqref{ratioE2} with fixed $\theta_{13}$ angles, $\theta_{13} = 8.2^\circ, 8.7^\circ, 9.1^\circ$, which is in $2\sigma$ range.

Actually, there exist solutions for our parameters $(A, B, C, a_2', a_3')$ in \eqref{MMN2} to realize these experimental values. For example, if we take the parameters to be 
\begin{equation}\label{solution}
\begin{aligned}
A &= 0.00198 \,\text{eV},  \\
B &= 30.5,  \\
C &= -5.94,  \\
a_2' &= -1.09,  \\
a_3' &= -1.06,
\end{aligned} 
\end{equation}
we can obtain $m_3 = 0.05\,\text{eV} , \theta_{13} = 8.7^\circ, \theta_{23} = 49.1^\circ$. This solution is also consistent with the $2\sigma$ range of recent fits from neutrinoless double beta decay \cite{Fogli:2012ua}:
\begin{align}
	m_{\beta\beta} &\approx 0.05 \text{\,eV},\\
	\Sigma &= m_1 + m_2 + m_3 \approx 0.15 \text{\,eV}.
\end{align}

\subsection{Charged lepton masses}
Next, we consider the charged lepton mass matrix \eqref{CLForU1E}. We want to fix the charged lepton masses as 
\begin{align}
m_e = \frac{y^e_l v_d}{\Lambda} \cdot d_1 &= 0.5 \times 10^6 \ {\rm eV}, \\
m_\mu = \frac{y^e_l v_d}{\Lambda} \cdot d_2 &= 105 \times 10^6 \ {\rm eV}, \\
m_\tau = \frac{y^e_l v_d}{\Lambda} \cdot d_3 &= 1776 \times 10^6 \ {\rm eV}.
\end{align}
The charged lepton masses are constrained from the D-flatness condition, which for this model is given by
\begin{align}
& \frac{7}{3} e_1 \vert d_1 \vert^2 + \frac{7}{3} e_2 \vert d_2 \vert^2 + \frac{7}{3} e_3 \vert d_3 \vert^2
- \frac{4}{3} e_1 \vert b_1 \vert^2 - \frac{4}{3} e_2 \vert b_2 \vert^2 - \frac{4}{3} e_3 \vert b_3 \vert^2 \nonumber \\
&+ \frac{2}{3} e_1 \vert a_1 \vert^2  + \frac{2}{3} e_2 \vert a_2 \vert^2  + \frac{2}{3} e_3 \vert a_3 \vert^2 
+ \frac{1}{3} e_1 \vert c_1 \vert^2 
= 0,
\end{align}
or equivalently
\begin{align} \label{dflat}
& + \frac{7}{3} \left( e_1 \vert m_e \vert^2 + e_2 \vert m_\mu \vert^2 + e_3 \vert m_\tau \vert^2 \right) \cdot \left\vert \frac{\Lambda}{y_1^e v_d}\right\vert^2 \nonumber\\
& - \frac{4}{3} \left( e_1 \vert a_2^\prime a_3^\prime A \vert^2 + e_2 \left\vert \frac{A  a_3^\prime}{a_2^\prime} \right\vert^2 + e_3 \left\vert \frac{A  a_2^\prime}{a_3^\prime} \right\vert^2 \right) \cdot \left\vert \frac{\lambda_1 \Lambda^2}{2\lambda_4 v_u^2 y_2^\nu} \right\vert^2\nonumber\\
&+ \frac{2}{3} \left(e_1 \vert A \vert^2  +  e_2 \vert a_2^\prime A \vert^2  + e_3 \vert a_3^\prime A \vert^2  \right) \cdot \left\vert \frac{ \Lambda^2}{v_u^2 y_2^\nu} \right\vert^2
+ \frac{1}{3} e_1 \vert A C \vert \cdot \left\vert \frac{\Lambda^3}{y_3^\nu v_u^2} \right\vert 
= 0.
\end{align}
After inserting the solution $(A, B, C, a_2', a_3' )$ from \eqref{solution} we can numerically solve \eqref{dflat} as a linear equation. Here, we only consider the simplified case where $\vert \lambda_1 / (2 \lambda_4) \vert = 1$. Then, we obtain a single solution,
\begin{align}
	\left\vert \frac{ y_1^e y_2^\nu}{y_3^\nu } v_d\right\vert & \approx 2.10 \text{\,GeV},\\ 
	\left\vert\frac{(y_1^e)^2}{y_3^\nu} \frac{v_d^2}{v_u^2} \Lambda \right\vert & \approx 7.06 \times 10^{12} \text{\, GeV}.
\end{align}
Then, by taking ``natural'' values, $\vert y_1^e \vert = \vert y_2^\nu \vert = \vert y_3^\nu \vert = 1$, and by imposing
\begin{align}
	v_u^2 + v_d^2 = (173 \text{\,GeV})^2
\end{align}
we arrive at
\begin{align}
	\tan \beta &= \frac{v_u}{v_d}  \approx 82.4,\\
	\Lambda & \approx 4.79 \times 10^{16} \text{\, GeV}.
\end{align}
Other values of $\tan \beta$ and $\Lambda$ are possible by appropriately adjusting the couplings.

\section{Conclusion}
In this work, motivated by a gauge origin of discrete symmetries in the framework of the heterotic orbifold models, we have investigated gauge theoretical realizations of non-Abelian discrete flavor symmetries.  We have shown that phenomenologically interesting discrete symmetries are realized effectively from a $U(1)^n \rtimes S_m$ or $U(1)^n \rtimes Z_m$ gauge theory. These theories can be regarded as UV completions of discrete flavor models. The main difference between a discrete flavor model and a $U(1)$ flavor model as shown in this paper can be seen in the field interactions. Namely, some fields in a discrete flavor model can be distinguished in a $U(1)$ flavor model. For example, the ${\bf 3}_{1(1)}$ representation field of the $\Delta(54)$ symmetry can be described by several $U(1)^2$ charges, $( e_1/3, e_2/3, e_3/3)$, $( -2e_1/3, -2e_2/3, -2e_3/3)$ etc. Thus a superpotential in a $U(1)$ flavor model can be different from the one of the corresponding discrete flavor model. In general, $U(1)^n \rtimes S_m$ and $U(1)^n \rtimes Z_m$ flavor models are constrained more than flavor models with non-Abelian discrete flavor symmetries, which are subgroups of $U(1)^n \rtimes S_m$ and $U(1)^n \rtimes Z_m$, because symmetries are larger. Our results would provide a new insight on flavor models.

We have introduced the specific combination of $U(1)^2$ charges, $e_1$, $e_2$, and $e_3$, to realize $S_4$, $\Delta(54)$, $A_4$ and $\Delta(27)$. They correspond to weights of the triplet (or anti-triplet) representation of $SU(3)$. In fact, $U(1)^2 \rtimes S_3$ is a subgroup of $SU(3)$, where $S_3$ is associated with the Weyl group. We also obtained genuine $U(1)^2 \rtimes S_3$ representations which are not obtained from $SU(3)$ triplets by spontaneous symmetry breaking. Also, in a stringy realization of $\Delta(54)$, the $SU(3)$ gauge symmetry appears in toroidal compactification, and the non-zero roots can be projected out by an orbifold projection \cite{Beye:2014nxa}. This may also suggest that a similar situation can be realized field-theoretically in a higher-dimensional $SU(3)$ gauge theory with a suitable orbifold boundary condition.

Anomalies of non-Abelian discrete symmetries are important \cite{Araki:2008ek}. Anomalous discrete symmetries would be violated by non-perturbative effects, but its breaking effects might be small depending on dynamical scales of non-perturbative effects. By our construction, discrete Abelian symmetries originating from $U(1)^n$ of $U(1)^n \rtimes S_m$ and $U(1)^n \rtimes Z_m$ are always anomaly-free and  exact symmetries, but  $S_m$ and $ Z_m$ of $U(1)^n \rtimes S_m$ and $U(1)^n \rtimes Z_m$ can include anomalous discrete symmetries depending on the model.
 
We have constructed a concrete flavor model for the lepton sector based on the $U(1)^2 \rtimes S_3$ continuous gauge theory. We have shown that it is possible obtain a realistic flavor structure from this model. Since the model is based on an extended symmetry the number of the parameters is relatively few. In particular, we could show a relation between the angle $\theta_{23}$ and third generation neutrino mass $m_3$. 

We have shown six types of gauge realizations of non-Abelian discrete symmetries. However, further extensions are possible. For example, extensions to higher $N$, $\Delta( 6 N^2 )$, is possible if we consider models with $U(1)$ charges $q = e_i/N$. It is also possible to include further representations of e.g. $U(1)^2 \rtimes S_3$ which we did not cover here for the sake of simplicity. The general representation theory of these semidirect groups is obtained from the little group method of Wigner, which is familiar from the representation theory of the Poincar\'e group. Then, e.g. in the case of $U(1)^2 \rtimes S_3$ one obtains an uncharged singlet representation which transforms as ${\bf 1}'$ under $S_3$ while being uncharged under the $U(1)^2$. 

A phenomenological implication of our $U(1)$ flavor models is that there should be $Z^\prime$ boson(s) which originate from $U(1)$ gauge groups in the effective theory. In this framework $Z^\prime$ bosons and flavor structures are related. Since we assigned different $U(1)$ charges to the three-generation leptons, the $Z^\prime$ bosons have flavor dependent interactions. Thus, if $Z^\prime$ bosons are light as e.g. the TeV scale, they can be a probe of the flavor structure. It will be interesting to investigate $Z^\prime$ phenomenology by extending well-known  discrete flavor models.


\subsection*{Acknowledgement}
S.K. wishes to thank Otto C.W. Kong for helpful discussions.
F.B. was supported by the Grant-in-Aid for Scientific Research from the Ministry of Education, Science, Sports, and Culture (MEXT), Japan (No. 23104011). T.K. was supported in part by the Grant-in-Aid for Scientific Research No.~25400252 from the Ministry of Education, Culture, Sports, Science and Technology of Japan. S.K. was supported by the Taiwan's National Science Council under grant  NSC102-2811-M-033-008.


\end{document}